\documentclass[11pt]{article}
\usepackage{amsfonts}
\usepackage{amsmath}
\usepackage{amssymb}
\usepackage{bbold}
\usepackage{geometry}
\usepackage{longtable}
\usepackage{graphicx}

\begin{document}
\begin{titlepage}
\vspace*{-1cm}
\phantom{hep-ph/***}
\flushright

\vskip 1.5cm
\begin{center}
\mathversion{bold}
{\LARGE\bf On the study of the Higgs properties at a muon  collider
}\\[3mm]
\mathversion{normal}
\vskip .3cm
\end{center}
\vskip 0.5  cm
\begin{center}
{\large Mario Greco}~$^{a)}$,
\\
\vskip .7cm
{\footnotesize
$^{a)}$~Dipartimento di Matematica e Fisica, Universit\`a  di Roma Tre;\\
INFN, Sezione di Roma Tre,\\
Via della Vasca Navale 84, I-00146 Rome, Italy
\vskip .5cm
\begin{minipage}[l]{.9\textwidth}
\begin{center} 
\textit{E-mail:} 
\tt{mario.greco@roma3.infn.it}
\end{center}
\end{minipage}
}
\end{center}
\vskip 1cm
\begin{abstract}

The discovery of the Higgs particle at 125 GeV is demanding a detailed knowledge of the properties of this fundamental component of 
the Standard Model.  To that aim various proposals of electron and muon colliders have been put forward for precision 
studies of the partial widths of the various decay channels. It is shown that in the case of a Higgs factory through a muon 
collider, sizeable  radiative effects - of order of 50\% - 
must be carefully taken into account for a precise measurement of the leptonic and total widths of the Higgs  particle. 
Similar effects do not apply in the case of Higgs production in electron-positron colliders. 
\end{abstract}
\end{titlepage}
\setcounter{footnote}{0}

\section{Introduction}
\label{intro}

The announcement of the discovery of a new scalar particle by the LHC experiments ATLAS \cite{Aad:2012tfa}  and CMS 
\cite{Chatrchyan:2012ufa} 
immediately has triggered the question of the real nature of this particle. The determination of  the spin 
- parity quantum numbers and the couplings to other Standard Model (SM) particles strongly
suggest it to be the Higgs boson, i.e. the particle responsible for the electroweak symmetry breaking . 
From the available data however, it cannot be concluded yet that we have found the SM Higgs boson and not one of the scalars 
postulated within the possible extensions of the SM. 

Therefore, a detailed study of this particle is required and  various types of Higgs factories have been proposed, 
which can precisely determine the properties of the Higgs boson, as an important  step in the future of high energy physics.
In particular the Higgs width's measurement is an essential ingredient to determine the partial width and the coupling constants to 
the fermions and bosons, and to this aim it is well known that  lepton colliders are most appropriate for precision measurements, 
as happened in the past for LEP and SLC.
Indeed the natural width  of a 125 GeV  SM Higgs boson is about 4 MeV, which is far from the precision which can be achieved at LHC.

Then circular electron-positron colliders are receiving  again considerable attention. Design studies have been launched at 
CERN with the Future Circular Colliders (FCC), formerly called TLEP \cite{Koratzinos:2013ncw}, of which an $e^{+}-e^{-}$ collider 
hosted in a tunnel of about 
100km is a potential first step (FCC-ee), and  in China with the Circular Electron Positron Collider (CEPC) \cite{wang}. 
Both projects can deliver very high luminosity    ($L > 10^{34}$ ) from the Z peak to HZ threshold. The Higgs bosons are
produced mainly through Higgs-strahlung, at center-of-mass energy of about 250 GeV, and therefore the Higgs boson is produced 
in association with a Z boson. Then the Higgs signal can be tagged with the Z boson decay, especially if the Z boson decays 
into a pair of leptons. The HZZ coupling can be inferred in this way, but it seems to us that the Higgs width is not achieved so easily.

On the other hand the idea of a muon collider with  $L > 10^{32}$ seems much more appealing (\cite{Cline:1993qpa},
\cite{Rubbia:2013iya}). 
The collider ring is much smaller, with $R \sim 50$ m, and here the direct $H^0$ cross section is greatly enhanced with respect
to $e^{+}-e^{-}$ since the s-channel coupling  is proportional to the square of the initial lepton mass. In analogy to the case of 
the $Z^0$ at LEP/SLC, the production of a single $H^0$ scalar in the s-state offers unique conditions of experimentation because 
the  $H^0$  has a very narrow width and most of its decay channels can be directly compared to the SM predictions with a very high 
accuracy The drawback of this project however  is that a very powerful "muon cooling" is needed \cite{Rubbia2}. 
A practical realization of a demonstrator of a muon cooled Higgs factory has been recently suggested by C. Rubbia \cite{Rubbia:2013iya}.
In the present note we show that in the case of a muon collider very important QED radiative corrections change drastically 
the lowest order Higgs production results, and must be taken in full account  for an accurate description of the Higgs total 
and partial widths, as well as the various  $H^0$ couplings. Such type of effects do not apply in the case of electron positron 
colliders, when the Higgs bosons are produced mainly through Higgs-strahlung in association with a Z boson.
The origin of these radiative effects is well known and has been discussed in very great detail in the case of collisions of 
electrons  and positrons with production of narrow resonances in the s channel like the J/Psi \cite{Greco:1975rm} and the 
Z boson \cite{Greco:1980mh}. 
Namely a correction factor $\propto \left(\Gamma / M \right)^{(4 \alpha / \pi)\log (2 E /M)}$ modifies the lowest order cross section, 
where M and $\Gamma$ are the mass and width of the s channel 
resonance, $W = 2 E$ is the total initial energy and m is the initial lepton mass. Physically this is understood by 
saying that the width provides a natural cut-off in damping the energy loss for radiation in the initial state.
To be more specific, defining 
\begin{eqnarray}
\beta_i = \frac{4 \alpha}{\pi} \left[\log \frac{W}{m_i}- \frac{1}{2}\right]\,,
\end{eqnarray}

where $m_i$ is the initial lepton mass,
\begin{eqnarray}
y &=& W - M \nonumber \\
\tan \delta_R(W) &=& \frac{1}{2} \Gamma / (-y) \nonumber
\end{eqnarray}

then the infrared factor $C^{res}_{infra}$ due to the soft radiation emitted from the initial charged leptons is given by [8]
\begin{eqnarray}
C^{res}_{infra} = \left(  \frac{y^2 + (\Gamma/2)^2}{( M/2)^2}\right)^{\beta_i/2} 
\left[1 + \beta_i \frac{y}{ \Gamma/2} \delta_R\right]\,,
\end{eqnarray}
   
so that the observed resonant cross section can be written as
\begin{eqnarray}
\sigma^c = C^{res}_{infra} \sigma_{res} (1 + C_F^{res})\,.
\end{eqnarray}
In the above eq. $\sigma_{res}$ is the Born resonant cross section (of Breit - Wigner form)  and $C_F^{res}$  is a finite standard 
correction of order $\alpha$ which we will neglect in the following.
In the case of Higgs production at a muon collider with W=2E = 125 GeV the factor  $\beta_i= 0.061$ and at the resonance (y = 0)  
the  factor  $C^{res}_{infra} =  \left(\Gamma / M\right)^{\beta_i}=0.53$, assuming the Higgs width  $\Gamma = 4$ MeV,  
which gives a substantial reduction of the Born cross section and therefore can mimic a smaller initial (and/or final)
partial decay width of the Higgs.

As it is well known, since the produced resonance is quite narrow, one has to integrate over the machine resolution, which is assumed to be
\begin{eqnarray}
G(W^\prime - W) = \frac{1}{\sqrt{2 \pi \sigma}} e^{-(W^\prime - W)^2/(2 \sigma^2)}
\end{eqnarray}
where  $\sigma$ is the machine dispersion, such that $(\Delta W)_{FWHM}=2.3548 \sigma$.
Then the experimentally observed cross section into a final state $| f >$ is given by
\begin{eqnarray}
\tilde \sigma(W) = \int G(W^\prime - W) dW^\prime \sigma(W^\prime)\,,
\end{eqnarray}
where
\begin{eqnarray}
\sigma(W^\prime) = \frac{4 \pi}{W^{\prime 2}} \frac{\Gamma_i \Gamma_f}{\Gamma^2} \sin^2 \delta_R (W^\prime)
\left\{\frac{\Gamma}{W^\prime \sin \delta_R (W^\prime)}\right\}^{\beta_i} \times 
\left(1 -\beta_i \delta_R \cot \delta_R \right) (1+C_F^{res})\,.
 \end{eqnarray}
 
By integrating last equation a useful formula for the observed cross section at the peak is  [8]
\begin{eqnarray}
\tilde \sigma(M) = \frac{2 \pi^2\Gamma_i \Gamma_f }{\sqrt{2} \pi \sigma M^2 \Gamma} 
\left(\frac{\Gamma}{M}\right) ^{\beta_i} e^{\left(\frac{\Gamma}{2 \sqrt{2} \sigma}\right)^2}
\left\{ercf \left(\frac{\Gamma}{2 \sqrt{2} \sigma}\right) + \frac{1}{2} \beta_i E_1 \left(\frac{\Gamma^2}{8 \sigma^2}\right)
\right\}(1 + C_F^{res})
\end{eqnarray}
where the second term in the square bracket represents the contribution from the radiative tail. 

Numerically, and neglecting the contribution from $C_F^{res}$, the overall radiative correction factor C  to the Born cross section 
at the peak is  C  =  0.47 ,   0.37 ,  0.30 ,  0.25  for  $\sigma$
 =  1 MeV,  2 MeV,  3 MeV,  4 MeV, respectively. This result shows once again the importance of the 
 radiative effects for a precision measurement of the Higgs  couplings.
 
To conclude, we have  shown that in the case of a Higgs factory through a muon collider, sizeable  radiative effects 
- of order of 50\% - must be carefully taken into account for a precise measurement of the leptonic and total widths of the 
Higgs  particle. Similar effects do not apply in the case of Higgs production in electron-positron colliders.

 

\end{document}